\documentclass[english,prl, twocolumn, aps]{revtex4-1}
\usepackage[T1]{fontenc}
\usepackage[latin9]{inputenc}
\usepackage{babel}
\usepackage{amstext}
\usepackage{graphicx}
\usepackage{esint}
\usepackage[unicode=true,pdfusetitle,
 bookmarks=true,bookmarksnumbered=false,bookmarksopen=false,
 breaklinks=false,pdfborder={0 0 1},backref=false,colorlinks=false]
 {hyperref}

\makeatletter
 
 \@ifundefined{textcolor}{}
 {%
   \definecolor{BLACK}{gray}{0}
   \definecolor{WHITE}{gray}{1}
   \definecolor{RED}{rgb}{1,0,0}
   \definecolor{GREEN}{rgb}{0,1,0}
   \definecolor{BLUE}{rgb}{0,0,1}
   \definecolor{CYAN}{cmyk}{1,0,0,0}
   \definecolor{MAGENTA}{cmyk}{0,1,0,0}
   \definecolor{YELLOW}{cmyk}{0,0,1,0}
 }

\makeatother

\begin{document}

\title{Green's function retrieval and fluctuations of cross density of states
in multiple scattering media}

\author{Julien de Rosny }

\email{julien.derosny@espci.fr}

\affiliation{Institut Langevin, ESPCI, CNRS, 1 rue Vauquelin, 75231 Paris cedex
05}

\author{Matthieu Davy}

\affiliation{Institut d\textquoteright{}Electronique et de Télécommunications
de Rennes, University of Rennes 1, Rennes 35042, France}
\begin{abstract}
In this article we derive the average and the variance of the cross-correlation
of a noise wavefield. The noise cross-correlation function (NCF) is
widely used to passively estimate the Green's function between two
probes and is proportional to the cross density of states (CDOS) in
photonic and plasmonic systems. We first explain from the ladder approximation
how the diffusion halo plays the role of secondary sources to reconstruct
the mean Green's function. We then show that fluctuations of NCF are
governed by several non-Gaussian correlations. An infinite-range NCF
correlation dominates CDOS fluctuations and proves that NCF is not
a self averaging quantity with respect to the plurality of noise sources.
The link between these correlations and the intensity ones is highlighted.
These results are supported by numerical simulations and are of importance
for passive imaging applications and material science.
\end{abstract}

\keywords{Correlations, multiple scattering, diffusion, local density of states,
Green's function retrial, seismology, material science, passive imaging}

\pacs{42.25.Dd, 42.30.Wb, 91.30.Dk }

\date{04/09/2013}

\maketitle
A wave propagating in a multiple scattering medium is completely scrambled
and generates random intensity patterns. Nevertheless, Weaver and
Lobkis \cite{Weaver2001} showed in 2001 that the time derivative
of the cross-correlation of an equipartitioned field measured at two
positions $\mathbf{r}_{A}$ and $\mathbf{r}_{B}$ is proportionnal
to the difference of the causal and anti-causal temporal Green's function
(GF). This result originates from the fluctuation-dissipation theorem
\cite{Callen1951} and has provided a framework for passive imaging
systems \cite{Wapenaar2006}. It has especially led to spectacular
developments in seismology where images of the earth crust have been
obtained at different scales with unprecedented resolutions \cite{Campillo2003,Shapiro2005}.
GF retrieval from cross-correlations of a diffuse field has also been
applied to acoustic waves \cite{Roux2004}, elastic waves \cite{weaver2003elastic,Wapenaar2004}
and recently electromagnetic waves \cite{Davy2013}. The NCF has been
interpreted as the field that is back-propagated by a time reversal
mirror that completely surrounds a multiple scattering medium \cite{Derode2003a}. 

In the frequency domain, the NCF function reduces to the imaginary
part of the GF, $\Im G(\mathbf{r}_{A},\mathbf{r}_{B})$. When the
positions of the probes coincide ($\mathbf{r}_{A}=\mathbf{r}_{B}$),
the NCF linearly depends on the local density of states (LDOS) which
counts the number of modes available at a given position. In optics,
the LDOS determines spontaneous and stimulated emission of light.
The LDOS exhibits spatial fluctuations caused by scatterers in the
vicinity of the source \cite{Mirlin2000,VanTiggelen2006,birowosuto2010observation,Caze2010}.
The variance of the LDOS is indeed equal to the intensity correlation
$C_{0}$ \cite{VanTiggelen2006,Caze2010}, which results from local
interaction. This infinite spatial range correlation was identified
by Shapiro \cite{Shapiro1999}. It differs from the universal intensity
correlations $C_{1}$, $C_{2}$ and $C_{3}$. The short-range contribution
$C_{1}$ simply results from Gaussian statistics. The non-Gaussian
contributions $C_{2}$ and $C_{3}$ are long- and infinite-range contributions,
respectively, and characterize statistics of enhanced intensity fluctuations
\cite{Akkermans2007}. 

About 10 years ago, Van Tiggelen \cite{Tiggelen2003} showed that
in a random medium, the NCF tends to be self-averaging even though
the noise sources are not equally distributed. Because multiple scattering
increases the spatial diversity of the field and therefore reduces
Gaussian fluctuations, the NCF converges more rapidly towards the
average NCF \cite{larose2006correlation}. The same conclusion was
derived from a parabolic approximation approach of scattering within
the time reversal framework \cite{Bal2002}. However, when the distance
between $\mathbf{r}_{A}$ and $\mathbf{r}_{B}$ is larger than one
elastic mean free path ($l_{e}$), the mean GF vanishes and the self-averaging
property of the NCF seems to be in contradiction with the deterministic
approach that claims that the NCF is given by $\Im G(\mathbf{r}_{A},\mathbf{r}_{B})$.
Indeed, even if $\left\Vert \mathbf{r}_{A}-\mathbf{r}_{B}\right\Vert \gg l_{e}$,
$\left\langle \left|\Im G(\mathbf{r}_{A},\mathbf{r}_{B})\right|^{2}\right\rangle >\text{0}$.
This result implies that non-Gaussian correlations should contribute
to NCF fluctuations. 

Here we first show that the ladder approximation helps to interpret
the emergence of the average NCF in a multiple scattering medium.
Scatterers located in one mean free path around the probes play the
role of secondary sources. Then we use a diagrammatic expansion of
the diffuse field to identify the significant non-Gaussian correlations
that characterize NCF fluctuations. We derive the analytical expression
of the variance of the NCF for one noise source or a continuous distribution
over the scattering volume. We show that the same infinite-range contribution
$\gamma_{2a}$ causes fluctuations of the cross density of states
(CDOS) and explains why the NCF is not self-averaging. We highlight
why this contribution cannot be deduced from the classical intensity
correlation term $C_{2}$. Finally, we shown that fluctuations of
the NCF in case of a single noise source are due to non universal
local terms. All those results are supported by numerical simulations.

We assume a set of uncorrelated wide-band sources of noise represented
by the power spectrum function $S_{V}(\mathbf{r})$ distributed over
a volume $V$ (or equivalently a surface in 2D). In the frequency
domain, the noise cross correlation $\zeta_{V}$ between two probes
at locations $\mathbf{r}_{A}$ and $\mathbf{r}_{B}$ is given by,

\begin{equation}
\zeta_{V}(\mathbf{r}_{A},\mathbf{r}_{B})=\int_{V}G^{*}(\mathbf{r}{}_{B},\mathbf{r})G(\mathbf{r}{}_{A},\mathbf{r})S_{V}(\mathbf{r})d^{d}\mathbf{r},\label{eq:integrale}
\end{equation}
where $d$ is the dimensionality of the space (here 2 or 3). The frequency
dependence is kept implicit. When the noise sources are uniformly
distributed ($S_{V}(r)=S_{\infty}$), the correlation of the fields
is integrated over the entire scattering volume and the NCF is proportional
to the imaginary part of the GF\cite{Weaver2004},

\begin{equation}
\zeta_{\infty}(\mathbf{r}_{A},\mathbf{r}_{B})=-\frac{l_{a}}{k_{0}}\Im G(\mathbf{r}_{A},\mathbf{r}_{B})S_{\infty}.\label{eq:ImG}
\end{equation}
Here $l_{a}$ is the absorption (inelastic) mean free path. The average
value $\left\langle \zeta_{V}\right\rangle $ is governed by$\left\langle G^{*}(\mathbf{r}{}_{2},\mathbf{r})G(\mathbf{r}{}_{1},\mathbf{r})\right\rangle $.
From the Bethe-Salpeter equation and the ladder approximation, $\left\langle \zeta_{V}\right\rangle $
is given by,
\begin{eqnarray}
\left\langle \zeta_{V}(\mathbf{r}{}_{A},\mathbf{r}_{B})\right\rangle  & = & \int\left\langle G(\mathbf{r}{}_{B},\mathbf{r})^{*}\right\rangle \left\langle G(\mathbf{r}{}_{A},\mathbf{r})\right\rangle S_{V}(\mathbf{r})d^{d}\mathbf{r}\nonumber \\
 &  & +\int\left\langle G(\mathbf{r}{}_{B},\mathbf{r}')^{*}\right\rangle \left\langle G(\mathbf{r}{}_{A},\mathbf{r}')\right\rangle F(\mathbf{r'})d^{d}\mathbf{r'}.\label{eq:meancorr}
\end{eqnarray}
The halo function $F(\mathbf{r}')$ is equal to $\int|\left\langle G(\mathbf{r},\mathbf{r}''\right\rangle |^{2}S_{V}(\mathbf{r})L(\mathbf{r}',\mathbf{r}'')d^{d}\mathbf{r''}d^{d}\mathbf{r}$.
The first term in Eq. (\ref{eq:meancorr}) is the coherent contribution
of the field. The second term can be interpreted using Eq. (\ref{eq:integrale}).
The expressions are indeed similar but the power spectrum function
is replaced by $F(\mathbf{r}')$ and the Green's functions are replaced
by the mean ones. The halo that diffuses from the noise source illuminates
the scatterers closer than an elastic mean free path from points A
and B. Those last scattering events therefore play the role of secondary
sources to build up the mean NCF. 

\begin{figure}
\includegraphics{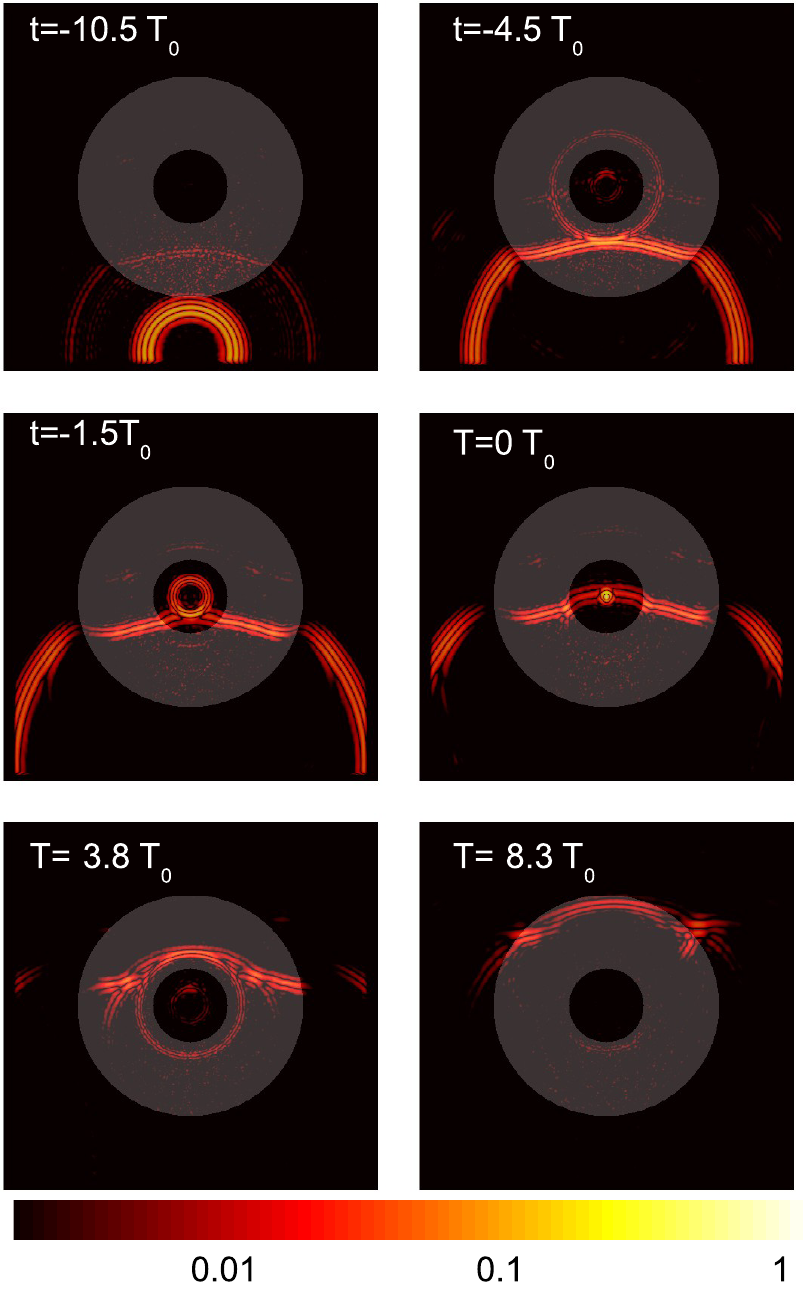}\caption{\label{fig:Average-cross-correlated-field}(a)-(f) Simulations of
the average cross-correlated field $\left\langle \zeta_{V}(\mathbf{r}_{A},\mathbf{r}_{B},t)\right\rangle $
for a single noise source shown at different times. The axis are the
coordinates of $\mathbf{r}_{B}$ with a 2D simulation domain of 50
by 50 wavelengths.The multiple scattering media is shown in gray.
The single noise source and the point A are located respectively at
the bottom of the scattering area and in the center of the figure. }
\end{figure}

To confirm this result, we carry out 2D numerical simulations in the
time domain with a finite-difference time-domain (FDTD) code. The
scatterers are uniformly distributed inside a ring with an inner radius
of $5\lambda_{0}$ ($\lambda_{0}$ is the central frequency wavelength)
and an outer radius of $20\lambda_{0}$. The mean free path is $\ell_{e}=1.5\lambda_{0}$
and the noise is emitted from a single source outside the multiple
scattering medium. The NCF $\langle\zeta_{V}(\mathbf{r}_{A},\mathbf{r}_{B},t)\rangle$
which is the inverse Fourier transform of Eq. (\ref{eq:meancorr})
is recorded and averaged over 270 disorder realizations. In Fig. \ref{fig:Average-cross-correlated-field}
the maps of the average NCF at different times are displayed versus
$\mathbf{r}_{B}$ for a fixed position $\mathbf{r}_{A}$. We clearly
observe an almost circular wavefront predicted by the coherent term
in (\ref{eq:meancorr}). A second circular wavefront focuses on point
A at negative times and is followed by a diverging one at positive
times. This contribution due to the halo appears with a skin depth
of about one mean free path. The result is even more spectacular on
an animation\cite{RosnyMovie}. Because the halo is almost uniformly
distributed over at least one mean free path around $\mathbf{r}{}_{A}$,
$\left\langle \zeta_{V}(\mathbf{r}{}_{A},\mathbf{r}_{B})\right\rangle $
is proportional to $\left\langle G^{*}(\mathbf{r}_{B},\mathbf{r}_{A})\right\rangle -\left\langle G(\mathbf{r}_{B},\mathbf{r}_{A})\right\rangle $
because $\int\left\langle G(\mathbf{r},\mathbf{r}_{A})\right\rangle \left\langle G^{*}(\mathbf{r},\mathbf{r}_{B})\right\rangle dV=-l_{e}/k_{0}\Im\left\langle G(\mathbf{r}_{B},\mathbf{r}_{A})\right\rangle $%
\footnote{This relation \cite{Akkermans2007} is similar to Eq. (\ref{eq:ImG})
but the elastic mean free path replaces the absorption mean free path
because the attenuation of mean field is mainly due to elastic scattering.%
} . The mean GF (resp. conjugate mean GF) represents the diverging
(resp. converging) coherent wave. 

In laboratory experiments, the NCF can easily be averaged over realizations
of the disorder. For instance, in optics the scatterers randomly move
as a consequence of the Brownian motion. In a microwave experiments,
the beads can be mixed in a toner. However, in seismology the NCF
can only be measured in a single realization of the disorder. The
NCF is expected to be self-averaging\cite{Bal2002,larose2006correlation}
as a result of Gaussian statistics. In the context of time reversal\cite{Derode2003}
or phase-conjugation focusing \cite{davy2012focusing}, long range
correlations that cannot be predicted by Gaussian fluctuations%
\footnote{The variance of a self-averaging quantity should fall towards 0 when
the quantity is integrated over time or space.%
} have been observed. We show in the following how those correlations
characterize the fluctuations of the NCF in disordered systems. To
this end, we estimated the variance $\gamma$ of $\zeta_{V}(\mathbf{r}_{A},\mathbf{r}_{B})$.
For simplicity, we have replaced $S_{V}$ by an integration over a
finite volume $V$ in Eq. (\ref{eq:integrale}). We first consider
an equipartitioned noise field, $V\rightarrow\infty$ and $\left\Vert \mathbf{r}_{A}-\mathbf{r}_{B}\right\Vert \gg l_{e}$.
Equation \ref{eq:ImG} gives, 
\begin{equation}
\gamma=\frac{l_{a}^{2}}{2k_{0}^{2}}\left\langle G(\mathbf{r}_{B},\mathbf{r}_{A})G^{*}(\mathbf{r}_{B},\mathbf{r}_{A})\right\rangle .\label{eq:gamma2a_final}
\end{equation}
This simple result shows that the NCF is not ``self-averaging''
in the sense that it does not converge towards the mean GF but towards
the \textit{exact} GF. The NCF is therefore sensitive to scatterings
that occur at a distance larger than a mean free path from the probes.
This fundamental result is crucial in monitoring applications. It
shows that it is possible to follow the evolution of a scatterer hidden
behind a multiple scattering media\cite{Bonneau2012}. 

\begin{figure}
\includegraphics{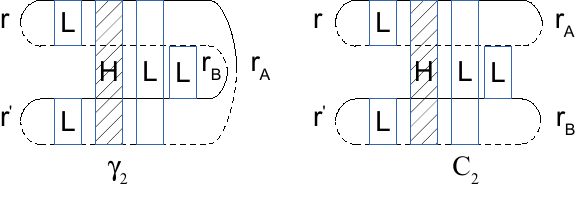}\caption{\label{fig:diagrams}Comparaison between the diagram that mainly contribute
to $\gamma_{2}$ and its C2 counterpart. More details can be found
in supplemental material \cite{rosnySup}. }
\end{figure}

In the following, we use the diagrammatic approach to interpret Eq.
(\ref{eq:gamma2a_final}) and to address fluctuations of the NCF for
a small number of sources. In the limit $\Delta r\gg\ell_{e}$, we
show in the supplemental material that the main contribution to $\gamma$
is $\gamma_{2a}$ shown in Fig. \ref{fig:diagrams}. Since the diagram
is long-range both in $\mathbf{r}$-$\mathbf{r}'$ and in $\mathbf{r}_{A}$-$\mathbf{r}_{B}$,
$\gamma_{2a}$ is of infinite range. In Fig. \ref{fig:diagrams} the
diagram of the long-range correlation $C_{2}$ widely used to characterize
intensity fluctuations is also depicted. Even though the two diagrams
look similar at first glance , we stress that $\gamma_{2a}$ is not
equal to $C_{2}$ because of an exchange between position $\mathbf{r}_{A}$
and $\mathbf{r}_{B}$ at the right side. $C_{2}$ is indeed short-range
in $\mathbf{r}_{A}$-$\mathbf{r}_{B}$.

We show in a supplemental material \cite{rosnySup} that the expression
of $\gamma_{2a}$ is

\begin{eqnarray}
\gamma_{2a} & = & 2\Delta^{4}\left(\Im\left\langle G(\mathbf{r},\mathbf{r})\right\rangle \right)^{2}h\left|\int_{V}L(\mathbf{r})d^{d}r\right|^{2}\nonumber \\
 &  & \times\frac{K}{D}L(\mathbf{r}_{B},\mathbf{r}_{A})\Im\left\langle G(\mathbf{r}_{A},\mathbf{r}_{A})\right\rangle \Im\left\langle G(\mathbf{r}_{B},\mathbf{r}_{B})\right\rangle .\label{eq:gamma2a}
\end{eqnarray}
Here $h$ is the Hikami constant and $\Delta=l_{e}/k_{0}$. The ladder
$L$ is solution of the steady state diffusion equation with absorption,
i.e, $-D\nabla^{2}L(\mathbf{r})+L(\mathbf{r})c/l_{a}=K\delta(\mathbf{r})$.
For 3D samples, $h=l_{e}^{5}/48\pi k^{2}$, $K=4\pi c/l_{e}{}^{2}$
and $D=l_{e}c/3$ and for 2D samples, $h=l_{e}^{5}/32k^{3}$, $K=4k_{0}/l_{e}$
and $D=l_{e}c/2$. In both cases, Eq. (\ref{eq:gamma2a}) yields $\gamma_{2a}=\frac{l_{a}^{2}}{2k_{0}^{2}}\left\langle G(\mathbf{r}_{B},\mathbf{r}_{A})G^{*}(\mathbf{r}_{B},\mathbf{r}_{A})\right\rangle $.
Since $\zeta_{\infty}(\mathbf{r}_{A},\mathbf{r}_{B})$ and CDOS are
both proportional to $\Im\left\langle G(\mathbf{r}_{A},\mathbf{r}_{B})\right\rangle $,
CDOS fluctuations also result from the same infinite-range term $\gamma_{2a}$.

The expansion of the Hikami vertex given in the SI that leads to Eq.
(\ref{eq:gamma2a}) is valid only for $\left\Vert \mathbf{r}_{A}-\mathbf{r}_{B}\right\Vert <l_{e}$.
Nevertheless in the case of coinciding probes $\mathbf{r}_{B}=\mathbf{r}_{A}$,
the NCF becomes proportional to the LDOS. NCF fluctuations are then
characterized by the correlation $\gamma_{0a}$ which is proportional
to the $C_{0}$ intensity correlation \cite{VanTiggelen2006,Caze2010}.
This non-gaussian term depends on the details of the local disorder
around the probe and involves a non-universal vertex $\chi_{0}$ \cite{Shapiro1999},
such as $\gamma_{0a}=2\delta V^{2}\Delta{}^{2}\left(\Im\left\langle G(\mathbf{r},\mathbf{r})\right\rangle \right)^{2}\left|\int_{V}L(\mathbf{r})d^{3}r\right|^{2}\chi_{0}$.
$\gamma_{0a}$ and $\gamma_{2a}$ can be seen as the two asymptotic
regimes $\Delta r\ll\ell_{e}$ and $\Delta r\gg\ell_{e}$, respectively,
of the variance $\gamma$ which characterizes the non-Gaussian fluctuations
of the NCF in mesoscopic multiple scattering media.

\begin{figure}
\includegraphics{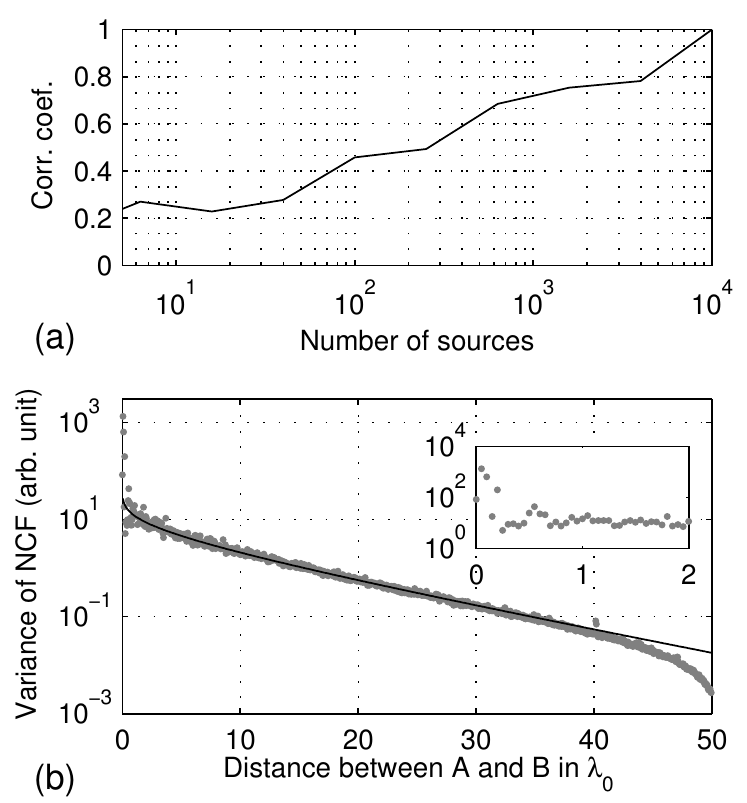}\caption{\label{fig:(a)amplitudemat} (a) Evolution with $N$ of the normalized
correlation coefficient between $\Im\left\langle G(\mathbf{r}_{A},\mathbf{r}_{B})\right\rangle $
and $\zeta$ for a single realization of disorder. (b) Variance of
$\zeta_{\infty}(\mathbf{r}_{A},\mathbf{r}_{B})$ ($N=10^{4}$) with
$\Delta r$ estimated from the averaging over 500 disorder realizations.
The black line is given by Eq.~(\ref{eq:gamma2a}). }
\end{figure}

We perform numerical simulations to support these derivations. The
2D multiple scattering medium is made of $10^{4}$ isotropic scatterers
enclosed in a disk of diameter 100$\lambda_{0}$. Here, $l_{a}\sim63\lambda_{0}$
and $l_{e}\sim2\lambda_{0}$. Those parameters ensure that the system
is in the diffusive regime. The sample is illuminated from $N$ independent
noise sources embedded in the medium. The NCF is computed from a scattering
matrix inversion method. For a single disorder realization, the NCF
is seen in Fig.~\ref{fig:(a)amplitudemat}(a) to converge towards
$\Im\left\langle G(\mathbf{r}_{A},\mathbf{r}_{B})\right\rangle $
for $N=10^{4}$. On Fig. \ref{fig:(a)amplitudemat}(b) we observe
that $\gamma$ is maximum for $\Delta r=0$ due to the $\gamma_{0a}$
contribution and then falls rapidly with $\Delta r$ until $\gamma_{0a}$
vanishes for $\Delta r\sim\lambda_{0}/2$. For $\Delta r\gg\ell_{e}$,
$\gamma$ decreases exponentially because of losses in the medium.
Simulations are in very good agreement with Eq. (\ref{eq:gamma2a})
in which the ladder $L$ is solution of the 2D diffusion equation
with losses, $L(\Delta r)=\beta K_{0}(\Delta r\sqrt{c/Dl_{a}})/2\pi D$.
This confirms that $\gamma_{2a}$ is of infinite-range.

\begin{figure}
\includegraphics{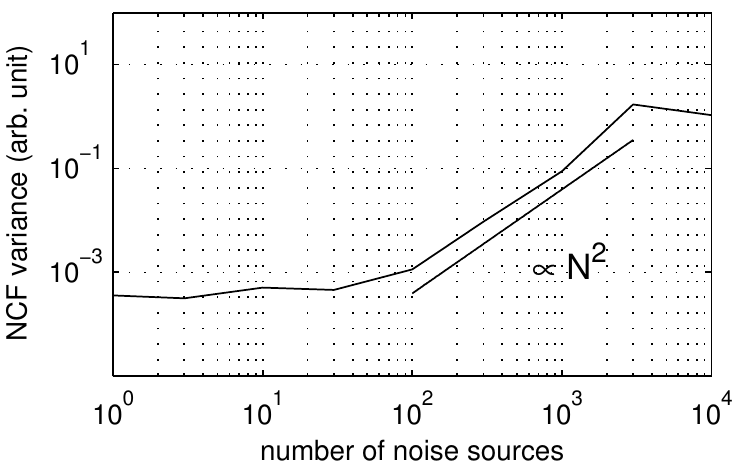}\caption{\label{fig:Continuous-cross-line}Simulations of the variance of the
NCF with $N$ for $\Delta r\sim15\ell_{e}$ obtained with 5e5 disorder
configurations. }
\end{figure}

The condition of an isotropic distribution of noise sources is most
of the time not fulfilled in seismology and in acoustics. For $\Delta r\gg\ell_{e}$,
$\gamma$ is seen in Fig. \ref{fig:Continuous-cross-line} to increase
as $N^{2}$ for $N>200$. In this range, the NCF is close to $\Im\left\langle G(\mathbf{r}_{A},\mathbf{r}_{B})\right\rangle $
in Fig. \ref{fig:(a)amplitudemat}(a) and the non-Gaussian contribution
$\gamma_{2a}$ (which scales as $N^{2}$) overcomes the $\gamma_{1}$
Gaussian contribution (C1-like contribution, see supplemental matetial)
which scales as $N$ \cite{larose2006correlation}. Finally, $\gamma$
saturates for $N\sim10^{4}$ since the sample is already illuminated
uniformly. Moreover $\gamma$ is larger than $\gamma_{1}$ for $N\sim1$.
This indicates that in addition to $\gamma_{1}$ other diagrams are
contributing to $\gamma$. Interference indeed occurs in the vicinity
of the noise source location $\mathbf{r}_{S}$ and another $C_{0}$-like
correlation has to be taken into account. Its expression is given
by $2\delta V^{2}\Delta^{2}L(\mathbf{r}_{B},\mathbf{r}_{S})L(\mathbf{r}_{A},\mathbf{r}_{S})\Im\left\langle G(\mathbf{r}_{A},\mathbf{r}_{A})\right\rangle \Im\left\langle G(\mathbf{r}_{B},\mathbf{r}_{B})\right\rangle \chi_{0}.$
The volume $\delta V$ of the single source is assumed smaller than
$l_{e}^{3}$. We finally note that in the case of $\Delta r\ll l_{e}$
and a single noise source, $\gamma$ is given by the sum of two $C_{0}$
contributions\cite{Retzker2002}, $\gamma=2\delta V^{2}\Delta^{2}\left|L(\mathbf{r}_{A},\mathbf{r}_{S})\right|^{2}\chi_{0}\left(\Im\left\langle G(\mathbf{r}_{A},\mathbf{r}_{A})\right\rangle ^{2}+\Im\left\langle G(\mathbf{r}_{S},\mathbf{r}_{S})\right\rangle ^{2}\right).$
Those considerations confirm the intuitive result that fluctuations
of NCF caused by a source located inside a multiple scattering medium
are stronger than fluctuations caused by a source outside this disordered
medium which only involves Gaussian fluctuations.

In conclusion, we have used the multiple scattering theory to demonstrate
the role of scatterers in the retrieval of the GF and to interpret
fluctuations of the NCF in terms of diffuse light interference. To
that end, we introduced an original diagrammatic contributions. Those
fundamental results can be applied to many different fields such as
seismology, acoustics, microwave, optics or material science. In acoustics
and in seismology, the estimation of the NCF is easily performed by
the direct cross-correlation of recorded time-depend fields. However,
the noise sources are usually not uniformly distributed and a generalization
of our approach to more complex source distributions would be a probe
of the convergence of the NCF towards the GF for a single realization
of disorder. This issue is of importance for imaging purposes. On
the other hand, in optics, one can take benefit of the thermal noise
that is uniform at thermal equilibrium. But then it is more tedious
to measure the NCF. In material science, metallic nanostructures can
for instance be excited with surface plasmons in disordered media.
Measuring the NCF at thermal equilibrium would make possible to estimate
the CDOS. We suggest the experiment consisting in the measurement
of the fluctuations of the field intensity diffracted by two tips
on a metallic surface at thermal equilibrium where plasmons are multiply
scattered to estimate $\gamma_{2}$ fluctuations. This would be an
extension of thermal radiation scanning tunneling microscopy\cite{DeWilde2006}. 
\begin{acknowledgments}
We wish to thank Roger Maynard, Philippe Roux, Remi Carminati and
Boris Shapiro for fruitful discussions. This work have been partially
supported by ANR grant ANR-10-BLAN-0124 OPTRANS.
\end{acknowledgments}

\end{document}